\RequirePackage{fix-cm}
\documentclass{article}
\usepackage[margin=22mm]{geometry}
\usepackage{setspace}
\usepackage{graphicx}
\usepackage{caption}
\usepackage[square, comma, sort&compress]{natbib}
\usepackage{amssymb}
\usepackage{algorithm}
\usepackage{algorithmic}
\usepackage{amsmath}
\usepackage{multicol}
\newcommand{\BigO}[1]{\ensuremath{\operatorname{O}\bigl(#1\bigr)}}
\begin{document}

\title{Forecasting and Event Detection in Internet Resource Dynamics using Time Series Models%\thanks{Grants or other notes
}

\author{S.P.Meenakshi\\ Department of Computer Science and Engineering \\ IIT Madras, India        \and
        S.V.Raghavan \\ Department of Computer Science and Engineering \\ IIT Madras, India %etc.
}


%%\institute{Indian Institute of Technology Madras \at
%%              Chennai -600036, India. \\
%%              Tel.: +914422575374\\
%%              \email{spmeena@cse.iitm.ac.in}           %  \\
%%           \and
%%            Indian Institute of Technology Madras \at
%%            Chennai -600036, India \\
%%            \email{svr@cse.iitm.ac.in}
%%}

\date{Received: \today / Accepted: date}

\maketitle

\begin{abstract}
 At present Internet has emerged as a country's predominant and viable data communication infrastructure.
 Autonomous System (AS) topology occupies the top  position in the Internet infrastructure hierarchy.
 The AS resources  which are building blocks of the
topology such as AS numbers, IPv4 and IPv6 Prefixes are globally competitive because of the finite resource pool.
 The resource requirement of each country is dynamic and driven by various technical and socio-economic factors.
 Hence the organizational and national competitiveness for socio-economic development is reflected in the
 AS growth pattern.
 Furthermore to assess the competitiveness, plan for future expansions
 and to make policies there is a need to study  and forecast the AS growth. As it is one of 
Internet infrastructure development 
indicators,
 understanding on long term trend and
 stochastic variation behaviour are essential to detect significant events during the growth period.
  In this work we  use time 
series based approximation 
for mathematical modelling, system identification and forecast the yearly AS growth.
 The AS data of five countries namely India, China, Japan, South Korea and Taiwan are extracted from APNIC
 archive for this purpose. The first two countries are larger economies and the next three countries are advanced 
technology nations in the APNIC region. The characterization of the time series is performed by analyzing
the trend and fluctuation component of the data. The model identification is carried out by testing for
 non stationarity and autocorrelation significance. ARIMA models with 
different Auto Regressive and Moving Average parameters are identified for forecasting the AS growth of each country.
 Model validation, parameter estimation, point forecast and prediction intervals with 95 $\%$ confidence
 levels for the five countries are reported in the paper. The statistical analysis on long term trends and 
change point detection on Inter Annual Absolute Variations (IAAV) are presented. The significant level change 
in variations, positive growth percentage in IAAV and higher percentage of advertised ASes when compared 
to other countries indicate India's fast growth  and wider global reachability of Internet infrastructure from 
2007 onwards. The correlation
 between AS IAAV  change point and GDP growth period indicates that the service sector industry growth 
is the driving force behind significant yearly changes.
 
\end{abstract}

\section{Introduction}
\label{intro}
Internet infrastructure plays a critical role in a country's economic, education and social development,
by increasing organizational and national competitiveness. In the rapidly growing communication infrastructure scenario,
Internet converges with other communication platforms such as public switched telecommunication networks and broadcast
medias. To better understand the Internet evolution, infrastructure and performance indicators are essential.
Infrastructure indicators  help to inform issues related to technological limitations and
 resource address portability,  level of
competition in the backbone market and convergence between different communication platforms.
 From a policy perspective too the indicators play a crucial role in
regulating the infrastructure.\\
\indent
 Autonomous System (AS) topology constitutes the top position in the Internet network hierarchy.
 The visibility of each network and their allocated IP addresses  to global data communication
 infrastructure is through AS numbers  and prefixes associated with the network.
Internet AS resources are globally competitive because of its finite resource
pool. The AS resource requirement of each country is dynamic and driven by various factors such as economic growth,
 industrial growth and government policies.  Hence to acquire the resources, plan for future expansions
 and make policies there is a need for forecasting. Also, as it is one of Internet infrastructure development 
indicators \citep{indicators},
 understanding on long term trend and stochastic variation behavior are essential to detect significant
 events during Internet growth.
\\
\indent
 The Internet resources are allocated by Internet Assigned Numbers Authority (IANA). The Regional Internet 
Registries (RIR) redistribute the resources directly to customers and  National Internet Registries   
(NIR). From the global AS pool, the RIRs  get working pool of 1024 blocks in order to meet
the current assignment demands. The AS resource requirements technically depend on number of large Internet
Service Providers (ISPs) deploying policy based services, multihoming users and usage of AS numbers as identifiers 
in MPLS VPNS \citep{ASN}. When a network uses different inter domain policies, then a public AS number is used
to realize each policy. In case of single policy network,  private AS numbers are used between the 
service providers and the BGP speaking client network. The reachability of the ASes and their prefixes 
are ensured by the global routing protocol E-BGP using routing data exchange process.
 Based on demand from different service providers and industries  the
AS resources are assigned from the working pool and maintained  by the RIRs. The AS numbers are unstructured
and have no direct aggregation advantage. But consecutive numbers can be used to separate effectively the domestic
and external traffic in firewalls implemented at Border ASes i.e., Great Firewall of China \citep{GFC}.
\\
\indent
One more technical concern in acquiring AS number is from 16 bit and 32 bit AS resource pool. 
Some of the existing AS domains routing E-BGP platform support only 16 bit AS numbers. Using 32 bit
AS number mean a transition of E-BGP platform to support 32 bit numbers for these domains. Since
BGP platform transition  takes place in its own temporal phase, 16-bit number AS numbers are
 most sought after until its predicted exhaustion in the year 2014. The 16 bit AS pool is almost exhausted.
 From the global 16 bit AS number pool of 64510, 61438 is already allocated 
to RIRs \cite{asn32}. For APNIC RIR, 16 bit AS numbers allocated from the global pool is 7828. In 
this 801 AS numbers are still unallocated as on 2013, February statistics obtained from the archive.
\\
\indent
The AS resource consumption of each country is dynamic. It is driven by various other factors such as economic growth,
 industrial growth and government policies apart from technical factors. In this work we  use time 
series based approximation for mathematical modeling, system identification and forecast the yearly AS growth.
 The AS data of five countries namely India, China, Japan, South Korea and Taiwan are extracted from APNIC
 archive for this study. The first two countries are larger economies and the next three countries are advanced 
technology nations in APNIC region. Considering the dynamics of resource consumption related to each country
appropriate model selection is required for forecasting.
\\
 \indent
The dynamics induced by factors like economic growth, increase in policy based ISPs, increased
multihoming users, competition for 16-bit AS resource pool and indirect summarization advantage have 
different impact on the long term growth trend and yearly variations. Also statistical analysis on AS data and 
the reason for temporal variations are sparsely addressed in the literature.  So in this work, we have attempted the
 statistical analysis on  country wise AS resource data with the following objectives: 
\begin{enumerate}
 \item To understand the country wise AS growth  that is influenced by normal and 
unusual occurrence of events, technology advancements and economic growth.
\item  To automate the process of  monitoring  and forecasting the temporal behaviors 
i.e, AS long term growth trend and yearly variations.
\item To understand the global reachability percentage of ASes and its growth.
\end{enumerate}
We accomplish our objectives by exploring the following questions.
\begin{enumerate}
 \item What are the characteristics of AS data of a country?
 \item Why ARIMA models are suitable to estimate the data for forecasting? 
 \item How the long term trend present for larger economies and technologically advanced countries?
 \item How to detect significant changes in yearly variations and account for it? 
 \item Why assigned Vs advertised AS number ratio different for the countries?
 \end{enumerate}
\indent
The contributions of the work are time series model identification, validation and forecasting AS growth. Long term
trend analysis, event detection for inter annual variations, significant event correlation with GDP and 
assigned Vs advertised AS comparisons for the countries under consideration.
\\
\indent
The  paper is organized as follows. The AS data analysis from APNIC data is discussed in section 2. 
AS data characterization, time series modeling and forecasting are discussed in section 3. Long term trend
analysis is reported in section 4. Inter annual variations and change point detections are discussed in section 5.
AS routeview data analysis is performed in section 6 followed by related work in section 7.
The conclusion of this work is presented in section 8. 
\section{AS Data analysis}
\label{sec:2}
The Internet resources: AS numbers, IPv4 prefixes and IPv6 prefixes are allocated and maintained by the RIRs
 such as APNIC, RIPE and ARIN. The ISPs, large network users such as content providers, corporates
and universities get the AS resources on request and statically  recorded in RIRs. The growth in allocated  AS 
resources of a country indicates the growth of the Internet infrastructure. We analyze the APNIC RIR data for 
the countries under consideration to understand the  growth behavior of these resources.
\subsection{APNIC AS Data}
The archived delegated resource data has been obtained from APNIC RIR. The registered AS number details related to
a country  are extracted from it. The  APNIC RIR  provides the allocated/assigned Internet 
resource statistics for the countries in Asia Pacific Region.
The resource statistics reported are for
\begin{enumerate}
 \item Autonomous Systems 
 \item IPv4 Addresses
 \item IPv6 Addresses
\end{enumerate}
 Public Internet address space and Autonomous System numbers allocated by Internet Assigned Numbers Authority (IANA)
 are redistributed by APNIC RIR to the countries in the region. The resources are managed with well structured policy
guidelines.
\\
\indent
We have used the record format details provided by the RIR to interpret the data 
\citep{APNIC}. The standard record format for the data entry in all the RIRs is as follows: 
$<$ registry, cc, type, start, value, date, status, extensions $>$.
Registry specifies the name of the RIR and cc indicates the country code to which the resource has been delegated.
The type field provides the details on whether the resource is AS number or IPv4 address or IPv6 address. The rest
of the fields give information on starting number, the total count of the resource from starting number, 
allocation/assigned date, whether
the resource is reserved or assigned to an organization and room for future extensions. 
 The aut-num attribute \citep{ASclass}
of AS class holds a short description or a name of the organization to which an AS number is assigned with. 
The summary report specifies
the total number of AS entries, IPv4 prefixes and IPv6 prefixes.
% The AS number entries related
%to the countries under consideration are extracted and given in Table \ref{table:APNIC_AS}. 
% The  actual and aggregated AS numbers observed on 01 January, 2012 and 2013 are placed in the 
%table.
\\
\section{Time Series Modeling}
\begin{figure}
 \centering
 \includegraphics[width=0.5\textwidth]{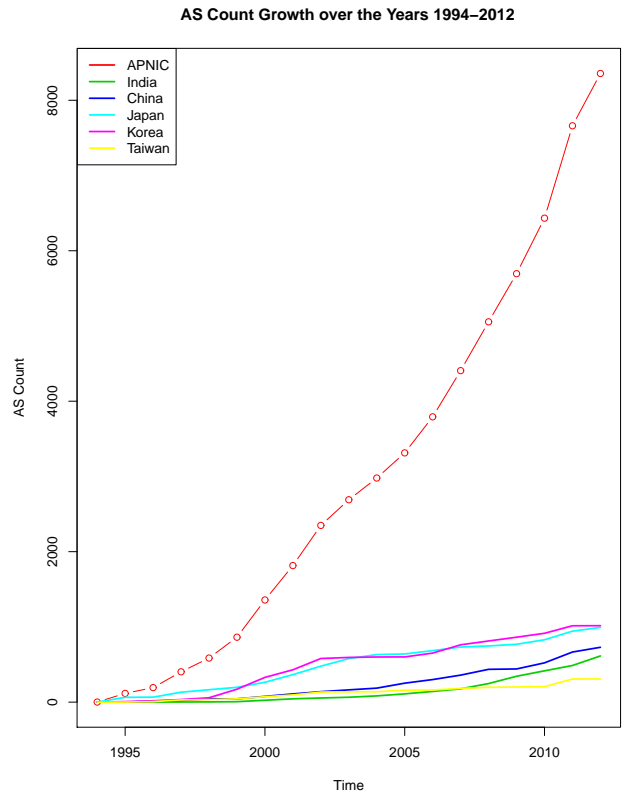}
 \caption{AS Count Growth Pattern Comparison}
 \label{fig:allsix}
\end{figure}

\begin{figure}
 \centering
 \includegraphics[width=0.5\textwidth]{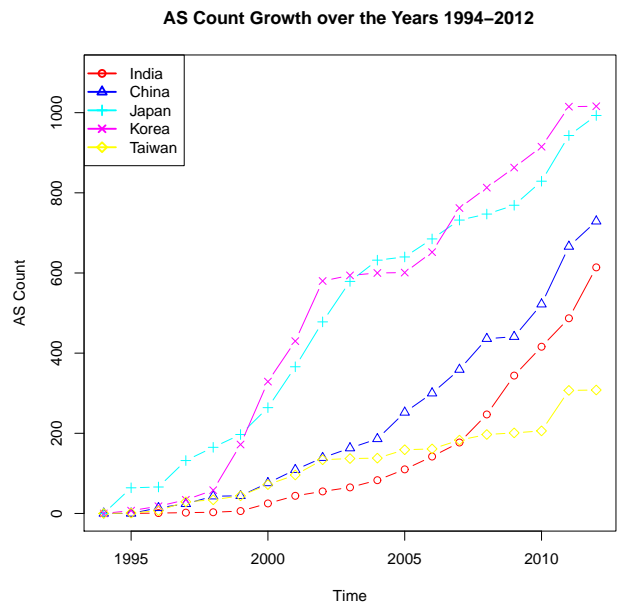}
 \caption{AS Count Growth Pattern for Five Countries}
 \label{fig:allfive}
\end{figure}

\begin{figure}
 \centering
 \includegraphics[width=0.5\textwidth]{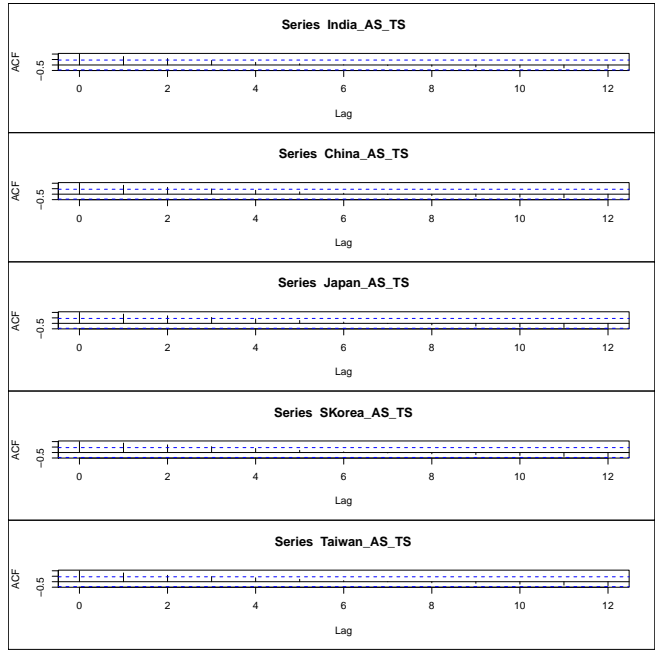}
 \caption{AS Year wise Growth ACF Plots}
 \label{fig:AS_ACF}
\end{figure}

\begin{figure}
 \centering
 \includegraphics[width=0.5\textwidth]{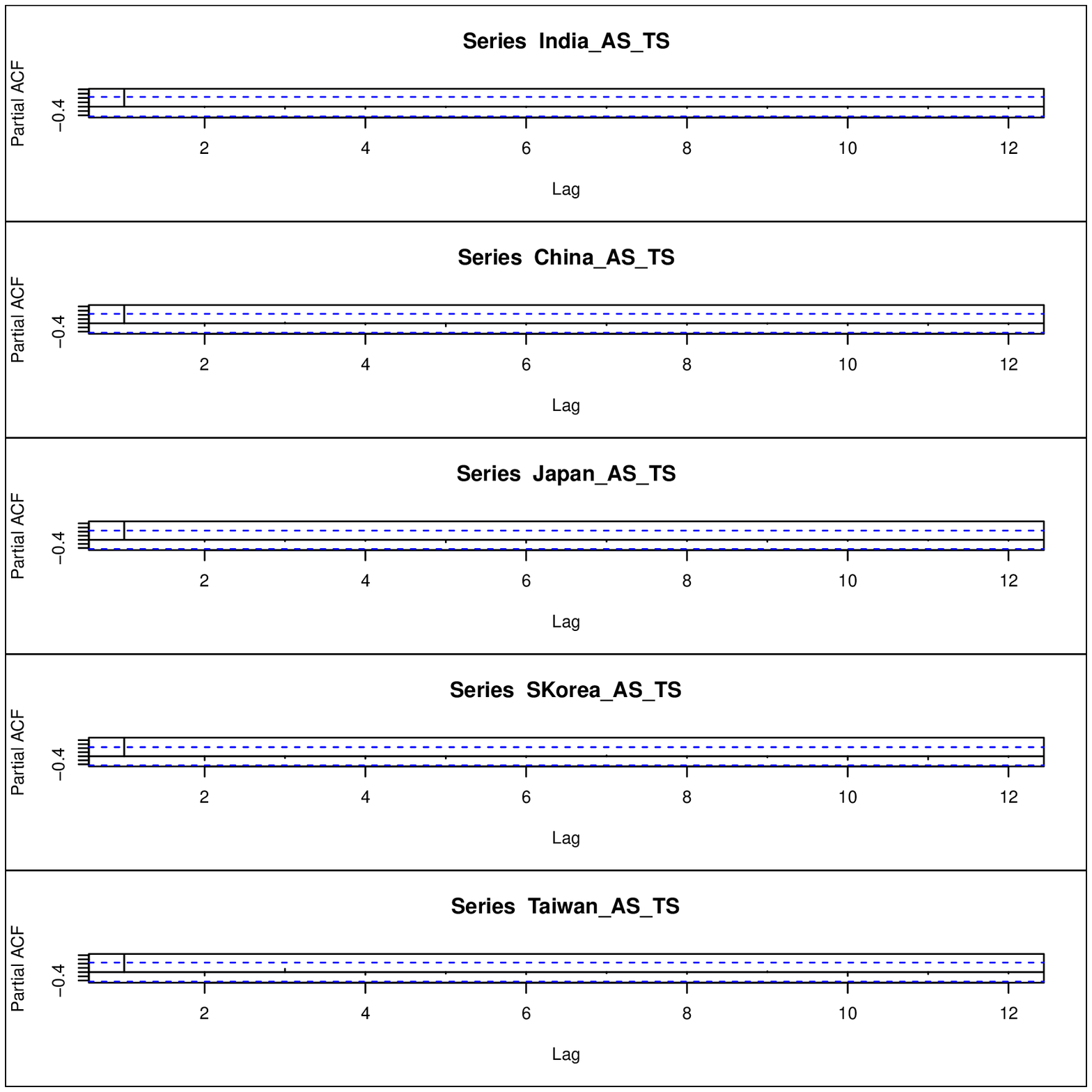}
 \caption{AS Year wise Growth PACF Plots}
 \label{fig:AS_PACF}
\end{figure}

\begin{figure}
 \centering
 \includegraphics[width=0.5\textwidth]{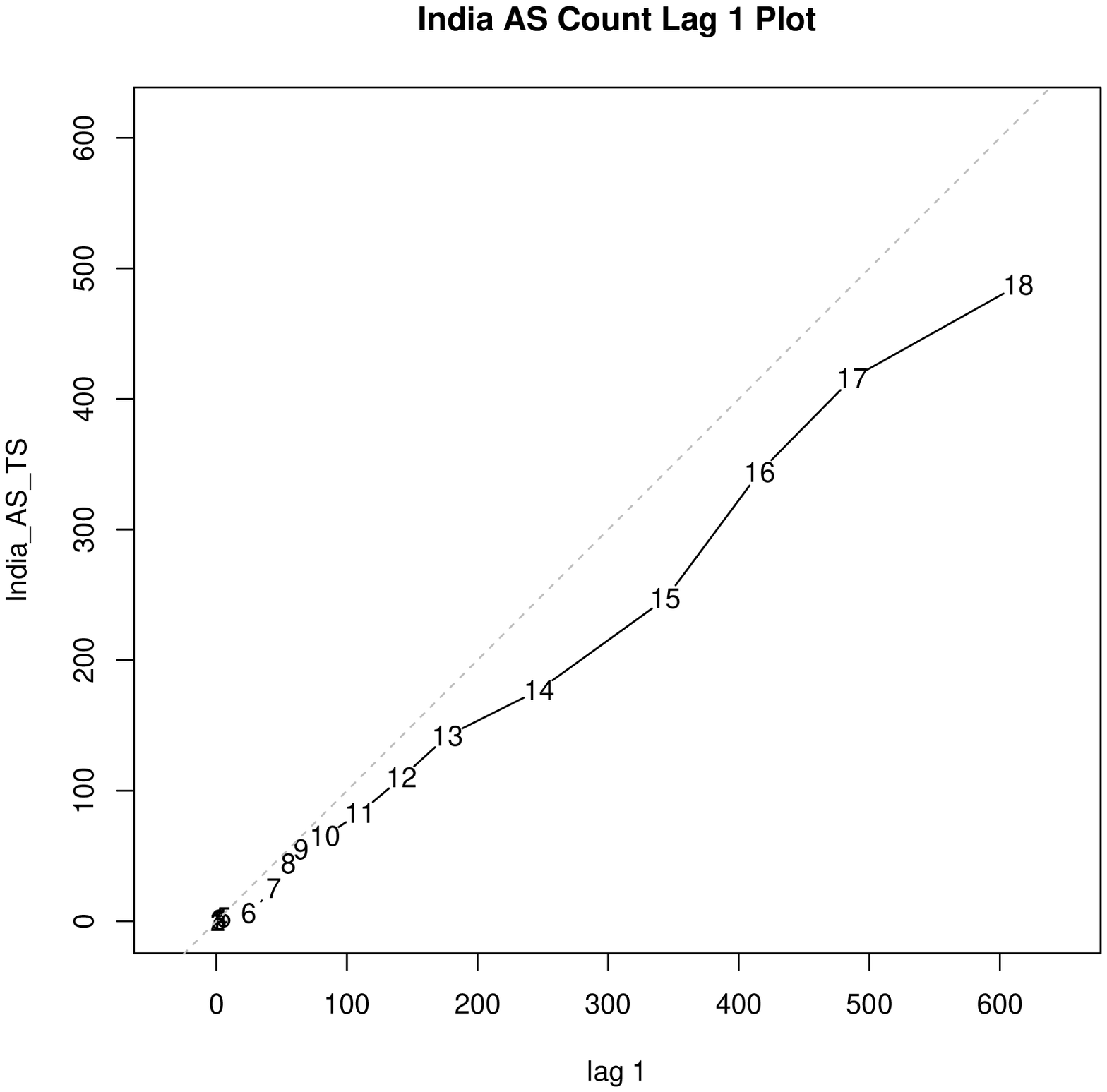}
 \caption{Lag 1 Plot - India}
 \label{fig:india_lag1}
\end{figure}

In our work, the AS count over the years is considered as non-stationary time series data. The non-stationary
property is confirmed using the Dickey Fuller hypothesis Test.
The data has been analyzed for
 modeling, forecasting  as well as to extract the following statistical properties and for detecting events.
\begin{enumerate}
 \item Long term trend on AS growth over the years
 \item Average Annual Growth rate and direction
 \item Inter Annual Absolute Variation (IAAV)  and direction
 \end{enumerate}
 The AS count 
time series for the APNIC region and the other five countries are given in Figures  \ref{fig:allsix} 
and \ref{fig:allfive}. 
\indent
We have the following observations on the AS count year wise data.
The AS number allocation in the APNIC region started in 1994. The countries Japan, South Korea and Taiwan
have  started their registrations from 1995 onwards which is an year ahead of India and China.
The year wise AS count of APNIC, India and China exhibits exponential growth trend. Japan South Korea and Taiwan
also exhibit exponential growth but combined with in between linear trends. As the data is annual data, there is no seasonality associated
with it. The variation is not constant and different variation levels are observed in the
growth trend. This property can't be captured using a single linear or non linear equation or
regression based models. This observation is also made in \citep{asmodel1,asmodel2}. Since the data is time series
, non stationary and the growth is influenced by many factors  such as economical growth of a country and
new content providers, we choose to use time series models. The time series model that captures the 
 trends of the data, relate the present values with past values and prediction error
is the ARIMA model. This model consists of Auto Regression (AR) component also known as lag term and 
Moving Average (MA) component also known as error term.
One of the significant criteria considered by us to select the best fit
 forecasting model among the model candidates is based on the accuracy of the forecasting 
and narrow prediction interval upto 5 years based on the observation. We also consider the other criterion such as 
minimum number of parameters, goodness of fit value and  residual independency with constant variance. 
The non seasonal general form of ARIMA model can be written using the equation \ref{equ:arima1}.
 \begin{equation}
 \label{equ:arima1}
 (1-\phi_1B - \cdots - \phi_p B^p)(1-B)^d (y_t - \mu t^d/d!) = (1 + \theta_1 B + \cdots + \theta_q B^q)e_t
 \end{equation}
where c = $\mu(1-\phi_1 - \cdots - \phi_p ) $ and $\mu$ is the mean of $(1-B)^d y_t$. $\phi_i$ is the AR coefficient 
and 'p' is the order of AR component. The MA coefficient is denoted by $\theta_i$ and the order of it is represented by
 'q'. The number
of required differencing is denoted by the term 'd'. The lagged terms of 'y' is specified by the notation 'B'. 
The error at time t is denoted by $e_t$.
 The specific values
for the order terms, coefficients and differencing terms are computed from the respective countries 
AS count data.
\\
\indent
The Auto Correlation Function (ACF) and Partial Auto Correlation Function (PACF) coefficients are 
computed between the series and its lag terms to identify the dependencies of the response variable
to its past observations. The order of AR and MA components are determined by analyzing the 
pattern of significant coefficient values present for the lag terms.
$Y_{1t},\dots,Y_{6t}$ represent the AS year wise growth of APNIC, India, China, 
 Japan, South Korea and Taiwan respectively. The ACF and PACF plots of AS year wise growth are given in the
Figures \ref{fig:AS_ACF} and \ref{fig:AS_PACF}. The significant values of the ACF and PACF indicate
 MA(2) and AR(1) order components for ARIMA model. The linear trend observed in Figure \ref{fig:allsix} 
supports for a first order differencing term in the  model. The linear dependency of the AS Count on its
lag 1 values shown in Figure \ref{fig:india_lag1} also suggests an Auto regressive model for the data.
 We select ARIMA(1,1,2), 
ARIMA(1,1,1) and  ARIMA(2,1,3) as candidate models for estimation to overcome the sample noise issue. 
 We present complete details on ARIMA  components estimation and validation for the time series of India. The
 forecasting results with prediction intervals are reported for all the countries under consideration.

\subsection{ARIMA Model and Residual Analysis}
\label{residual}
\begin{figure}
 \centering
 \includegraphics[width=0.5\textwidth]{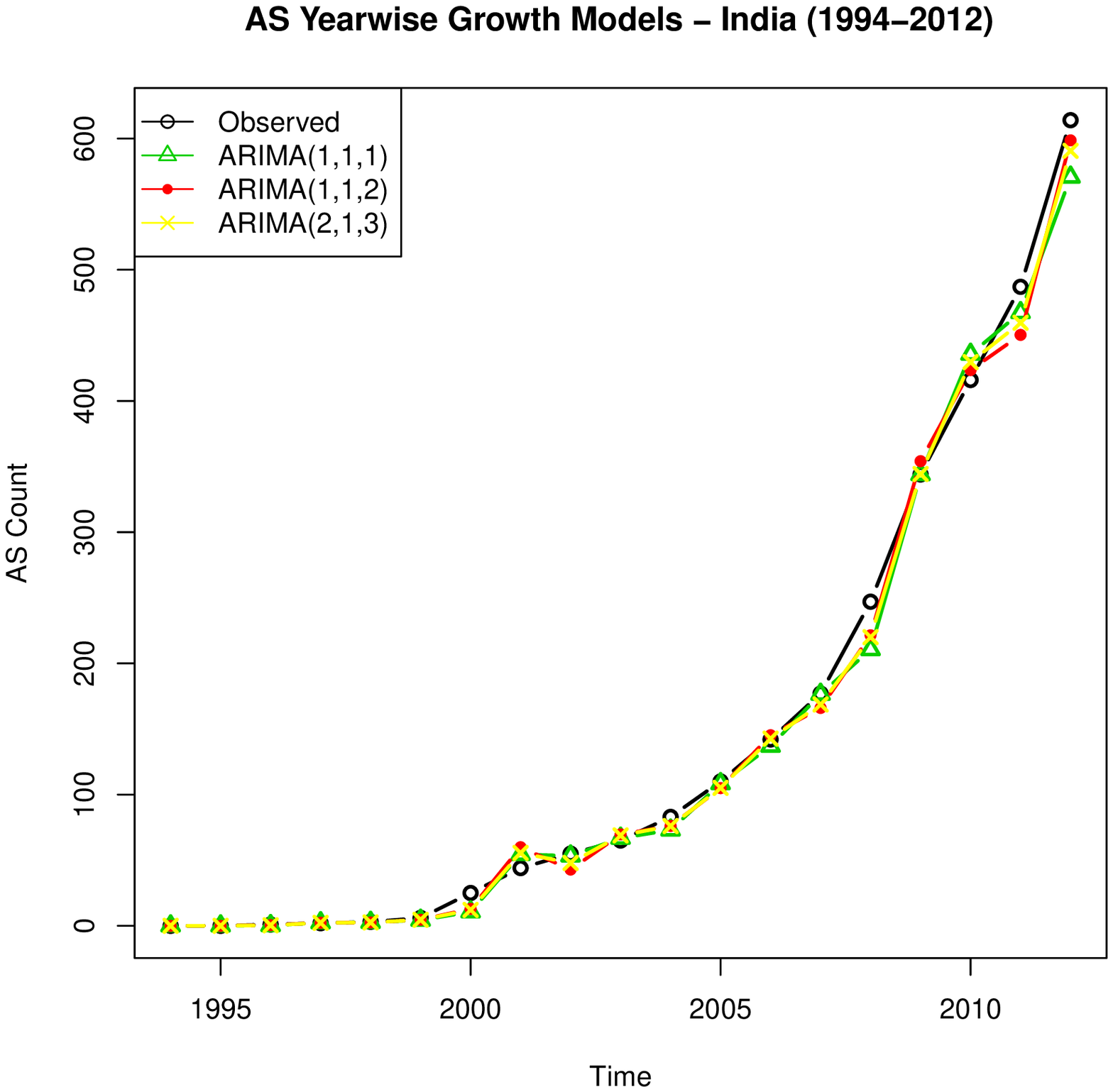}
 \caption{AS Year wise Growth Models - India}
 \label{fig:AS_models}
\end{figure}

\begin{figure}
 \centering
 \includegraphics[width=0.5\textwidth]{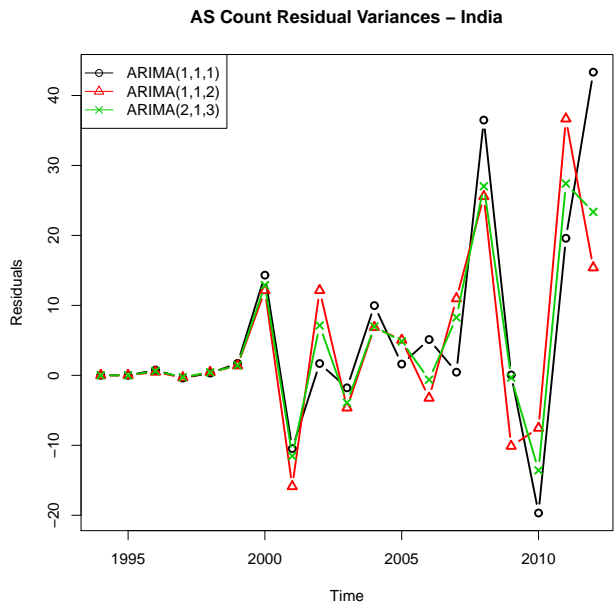}
 \caption{Model Residual Variances - India }
 \label{fig:as_res_var1}
\end{figure}

\begin{figure}
 \centering
 \includegraphics[width=0.5\textwidth]{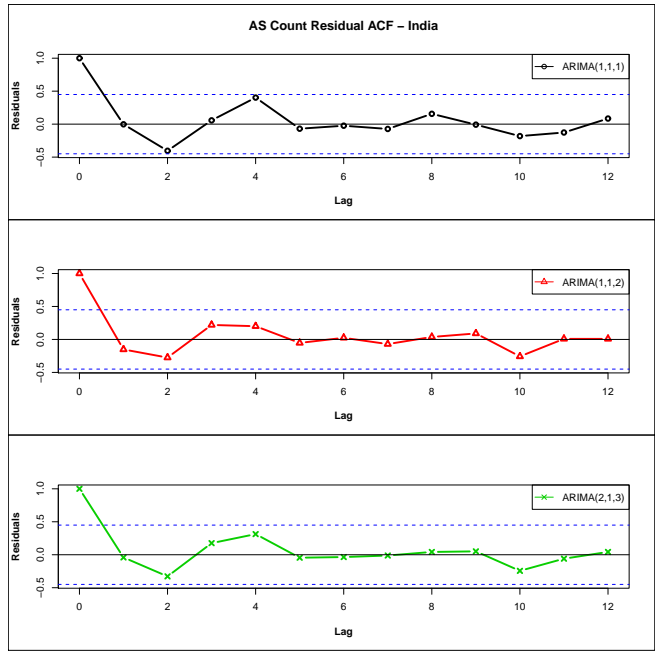}
 \caption{Model Residual ACF - India }
 \label{fig:as_res_acf1}
\end{figure}

The AS year wise growth data is modeled using the candidate ARIMA models. The regression is performed
with observed predictor values on the response variables using minimization of Conditional Sum of Squares (CSS)
 method combined with maximum likelihood (ML) method. The non seasonal
ordering is used in the model. ARIMA models do internal mathematical transformations on the series
 to detrend and stabilize the variance of the data. The transformations on data for detrending include
differencing. For variance stabilization, functions such as log and square roots are used.
The predictions are performed on the transformed data and been converted to original
series by reversing the transformation process. We have used the R statistical package for modeling \citep{Rmodel,Rmodel1}.  
The fitted models to the Indian AS year wise data are shown in Figure \ref{fig:AS_models}. The ARIMA(1,1,2) visual
fitness is good for the data.
The estimated model component coefficients ($\phi_i$,$\theta_i$), standard error, z-statistic and AICc are given in table \ref{table:AS_details}. 
The z-statistic  is computed as ratio between component coefficient and standard error. The parameter is statistically
 significant when z statistic is greater than 1.96. AICc is a goodness of fit measure computed based on number of parameters
 and information loss. The ARIMA(2,1,3) model coefficients namely AR2, MA2 and MA3  are not 
statistically significant since z statistic is less than 1.96. Also AICc increases and variance decreases
 along with the number of
parameters. This may be accounted to over fitting of the observed values and
 more information loss caused by the increase in number of parameters of the models. 
\begin{table}
\caption{ARIMA Model Estimation for AS Year wise Growth Data - India}
\label{table:AS_details}       % Give a unique label
\begin{center}
\begin{tabular}{|c|c|c|c|c|c|c|}
\hline\noalign{\smallskip}
 Model & AR Coeff.$\phi_i$, & z & MA Coeff.$\theta_i$, & z & AICc & Sample\\
       & Std. error&  Statistic &Std error &Statistic & &Variance\\
\noalign{\smallskip}\hline\noalign{\smallskip}
\hline
 ARIMA(1,1,1)& (0.96,0.07) & 13.71  & (0.85,0.16) & 5.3 & 163 &246\\
 \hline
 ARIMA(1,1,2)& (0.76,0.18) & 4.22 & (1.74,0.28)&6.21&164&176 \\
             &             &      & (1.0,0.31)&3.2 &  &\\
 \hline     
 ARIMA(2,1,3)& (0.87,0.40) & 2.18 &(1.26,0.37) & 3.4 & 169&151 \\
           & (0.099,0.388)& 0.25& (0.057,0.454)&0.125&&\\
           &               &       & (-0.53,0.28)& -1.89&&\\
\noalign{\smallskip}\hline

 \end{tabular}
\end{center}

\end{table}
\\
\indent
Residuals are analyzed to validate the fitness of the model for the observed data.
The variances of the residuals for all the three models are shown in Figure \ref{fig:as_res_var1}. 
The time series plot of the residuals does not exhibit any abnormalities and the variances are roughly constant 
over a period of time.
The residual ACF given in Figure \ref{fig:as_res_acf1} for all the three models have no significant evidence for 
autocorrelation upto lag 12. Hypothesis tests for checking white noise property of the residuals are
done using Jarque Bera Test and Shapiro-Wilk normality test. The null hypothesis for both the tests is 
that the residual series follow a normal distribution. The p-values for each model is 
given in table \ref{table:AS_res_nor}. ARIMA(1,1,2) model has p -value greater than 0.05 for both the tests and hence 
null hypothesis is accepted. The other two models fail in  both or any one of the tests.
\begin{table}
\caption{AS Count Model Residual Normality Test Results}
\label{table:AS_res_nor}       % Give a unique label
\begin{center}
\begin{tabular}{|c|c|c|}
\hline\noalign{\smallskip}
 Model & Jarque Bera &  Shapiro-Wilk \\
       & p-value   &   p-value \\
\noalign{\smallskip}\hline\noalign{\smallskip}
\hline
 ARIMA(1,1,1)&0.04362  & 0.002736 \\
 \hline
 ARIMA(1,1,2)& 0.2215 & 0.2248  \\
\hline     
 ARIMA(2,1,3)& 0.4522 & 0.03706 \\
           
\noalign{\smallskip}\hline
 \end{tabular}
\end{center}
\end{table}
\subsection{Forecasting}
\begin{figure}
 \centering
 \includegraphics[width=0.5\textwidth]{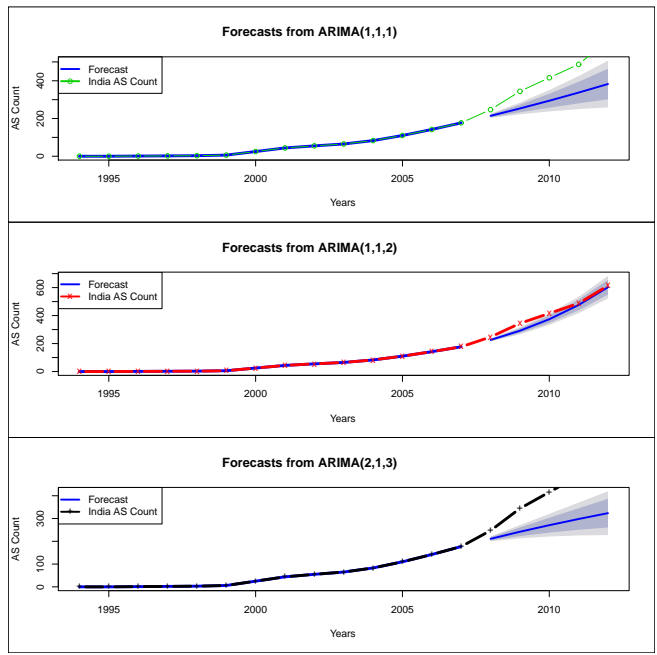}
 \caption{Model Forecasts With Prediction Intervals - India }
 \label{fig:AS_fore1}
\end{figure}

Forecasting of AS year wise growth data is computed using the three models. We have taken 14 values in the 19 annual data
for estimating each model and predict the next 5 year values. Point forecasts as well as prediction intervals are
computed. The prediction interval is computed using the standard error of the forecasts. The 95 percentage confidence
interval is computed by assuming the model error follows normal distribution. Forecast standard error 
is computed from Psi weights ({$\psi_i$}) that are obtained by converting the ARIMA model to infinite order MA model.
 The prediction interval is computed as current predicted value
$\pm$ 1.96 times the standard forecast error. The forecast series are shown in Figure \ref{fig:AS_fore1}.
\\
\indent
From the graphical analysis, it is observed that ARIMA(1,1,2) model fits and forecasts the year wise AS data within the 
prediction intervals. For the other two models, the forecast values are well beyond the prediction interval except
the first forecast value. Also the prediction interval is narrow for the ARIMA(1,1,2) model which indicates
that variability of the data is stabilized in the model. We have given the 
point forecasts, 95 $\%$ confidence level prediction interval and Root Mean Square Error (RMSE) accuracy 
in the table \ref{table:AS_fore1}. The accuracy is computed as square root of mean squared errors which occur
 between the predicted value and the observed value. ARIMA(1,1,2) model has good point forecast, prediction interval 
and RMSE accuracy when compared to other models to the given observations. The ACF and PACF analysis,
 visual fitness of the model, number of parameter significance, residual analysis and prediction properties confirms
the adequacy of ARIMA(1,1,2) model for the AS count year wise growth data.
 \begin{table}
\caption{AS Count Year wise Growth Model Forecasting -India}
\label{table:AS_fore1}  
     % Give a unique label
\begin{center}
\begin{tabular}{|c|c|c|c|}
\hline\noalign{\smallskip}
 Model & Data &  Prediction  & Accuracy\\
       & Predictions &  Interval &  RMSE\\
\noalign{\smallskip}\hline\noalign{\smallskip}
\hline
Observed    &(247,344,416,487,614)&&\\
\hline
 ARIMA(1,1,1)&(214,253,294,337,383)&(203-225),(222-284),& 141.36 \\
             &                     & (238-351),( 250-425),&\\
             &                     & (259-507) & \\
\hline
 ARIMA(1,1,2)&(226,292,374,476,603) &(215-236),(266-317),& 32.32 \\
             &                      &(333-414),(417-534),&\\
             &                      &(522-683)           &\\
\hline       
 ARIMA(2,1,3)&(211,242,271,298,324) &(200-222),(213-271),& 174.9\\
             &                      &(221-320),(226-370),&\\
             &                      & (228-420)          & \\
\noalign{\smallskip}\hline
 \end{tabular}
\end{center}
\end{table}
\\
\indent
Similar procedure is followed for model estimation, validation, forecasting and selection to the other four countries 
and APNIC region. For the APNIC region, when compared to other models ARIMA(1,1,1) estimation is visually good.
 The MA1 coefficient is less than the z - statistic threshold. Variances are roughly constant and no evidence 
of autocorrelation is found upto lag 12. Both the normality tests confirm the normal distribution of the residuals.
 RMSE is also less and most of the forecast values fall within 95 $\%$ confidence interval. Similar
observations are made to this model for the countries Japan, South Korea and Taiwan.
For China, when compared to other models ARIMA(2,1,3) is relatively good with the expected  residual and forecasting
properties. The parameters and forecast observations for the selected models are given in the tables
\ref{table:AS_paramAll} and \ref{table:AS_foreAll}.
 \begin{table}
\caption{ Selected Model Estimations for AS Year Wise Growth Data }
\label{table:AS_paramAll}       % Give a unique label
\begin{center}
\begin{tabular}{|c|c|c|c|}
\hline\noalign{\smallskip}
 Region / &Model & AR (Parameter,  & MA (Parameter,  \\
 Country   &      &      Std. error)    &  Std error)\\
\noalign{\smallskip}\hline\noalign{\smallskip}
\hline
 APNIC& ARIMA(1,1,1)&(0.975,0.038)&(-0.33,0.24)\\
\hline
 India& ARIMA(1,1,2)&(0.76,0.18)& (1.74,0.28) \\
      &             &           & (9.00,0.31)\\
\hline
 China& ARIMA(2,1,3)&(0.91,0.27) & (-0.62,0.47)\\
      &             & (0.061,0.269)& (-0.62,0.67)\\
      &              &              & (1.00,0.48)\\
\hline 
 Japan& ARIMA(1,1,1)& (0.91,0.14) & (-0.33,0.44) \\
 \hline
 SKorea& ARIMA(1,1,1)& (0.75,0.20) & (-0.10,0.32)\\
\hline
Taiwan & ARIMA(1,1,1)& (1.0,.005)& (-0.98,0.12)\\
   \noalign{\smallskip}\hline
 \end{tabular}
\end{center}
\end{table}

\begin{table}
\caption{Selected Model Forecasting for AS Year wise Growth Data }
\label{table:AS_foreAll}       % Give a unique label
\begin{center}
\begin{tabular}{|c|c|c|c|c|c|}
\hline\noalign{\smallskip}
 Region / & Model & Observation  & Point & Prediction & Accuracy    \\
 Country  &       &              &  Forecast  & Interval   & RMSE        \\
\noalign{\smallskip}\hline\noalign{\smallskip}
\hline
 APNIC& ARIMA(1,1,1)&(5055,5695,&(5064,5765,&(4839-5288,5243-6287,& 185.52\\
      &            & 6433,7661 & 6513,7311, &5613-7413,5955-8666&\\
      &            & 8356)     &  8161)      & 6275-10048)&\\
\hline
 India& ARIMA(1,1,2)&(247, 344 &(226,292, & (215-236,266-317, & 32.32\\  
      &             &416, 487 & 374,476,  & 333-414,417-534,  & \\
      &             & 614)   & 603)      & 522-683)          & \\
\hline
 China& ARIMA(2,1,3)&(436,441,&(391,434,& (365-417,381-488, & 115.48\\
      &             & 522,666)&469,506, &  377-560,382-629, &\\
      &             & 729)    & 539)    &   381-698)        &\\ 
\hline 
 Japan& ARIMA(1,1,1)&(747,769& (771,805 & (701-841,666-944, & 64.42 \\
      &             & 829,943& 835,861  &  620-1049,567-1154&     \\ 
      &             & 993)   & 883)     &  507-1259) & \\
\hline
 SKorea& ARIMA(1,1,1)&(813,863&(851,924 &(742-961,696-1151,& 49.64 \\
       &             &915,1015 &982,1029 &628-1335,545-1512,& \\
       &             & 1016)   & 1067  )& 454-1679)&\\
\hline
Taiwan & ARIMA(1,1,1)&(197,201,&(197,211, &(170-225,166-257 & 39.89  \\
       &             & 206,307 & 226,240, & 163-289,158-321 &\\   
       &             & 308)    & 254 )    & 153-355)   &\\
 \noalign{\smallskip}\hline
 \end{tabular}
\end{center}
\end{table}
\section{Long Term Trend Analysis }
\label{sec:3}
\begin{figure}
 \centering
 \includegraphics[width=0.5\textwidth]{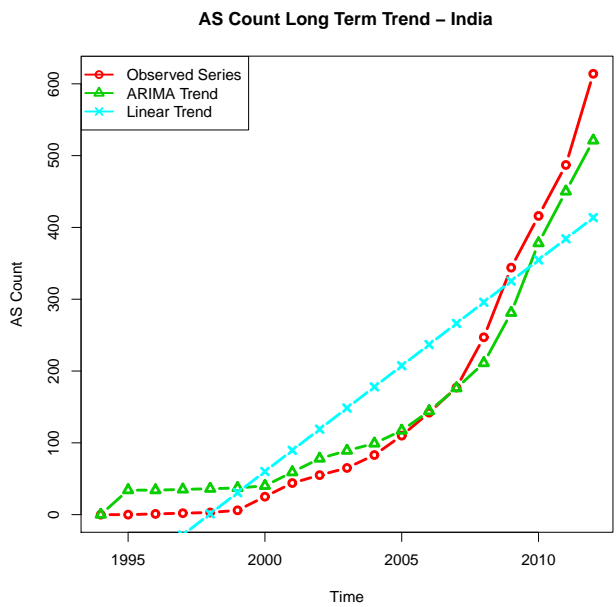}
 \caption{AS Count Long Term Trend (1996-2012) - India}
 \label{fig:AS_ltrend1}
\end{figure}

\begin{figure}
 \centering
 \includegraphics[width=0.5\textwidth]{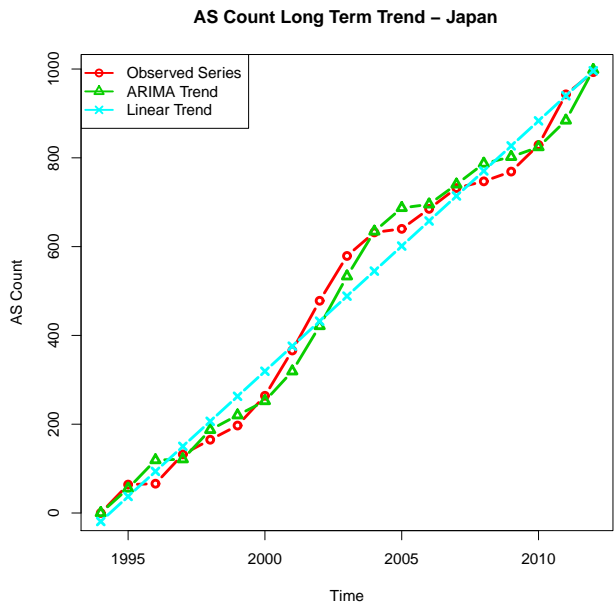}
 \caption{AS Count Long Term Trend (1995 -2012) - Japan}
 \label{fig:AS_ltrend3}
\end{figure}

The long term trend which is a component of AS data time series is estimated and analyzed in this section.
As we observed before, AS data has additive long term trend. This trend can be estimated from
the observations using a statistical model that explicitly includes local or global trend parameter.
We choose random walk model with drift (ARIMA(0,1,0)) and linear model to estimate the trend.
The drift parameter of ARIMA(0,1,0) and  slope of the linear model represent the  average 
yearly growth rate. In addition to that the slope of the linear model provides the direction of the growth.
The estimated growth trend using both the models to India and Japan are shown in Figures \ref{fig:AS_ltrend1}
and \ref{fig:AS_ltrend3}. From the long term trend graph we have observed that the actual AS count observations
 cycles around the linear trend line. The half cycle period is approximately 12 years for APNIC, India and China.
 But for technology advanced countries the half cycle period is approximately 5 years.
\indent
The table \ref{table:AS_ltrend} gives the trend values for all the countries under consideration. Since the average
 growth rate computed by both the methods are almost similar, we take average from these two methods for further analysis.
Technology advanced countries Japan, Korea and Taiwan have average annual growth rate of 56, 62 and 18 AS counts 
respectively.
Among larger economies India and China  have average annual growth rate of 36 and 44 AS counts.
The APNIC region has annual average growth rate of 459 AS counts. Relative average growth rate with
respect to APNIC region is computed for comparison purpose. India is
1.8  $\%$ less in relative average growth than China. Approximately 5  $\%$ less in average growth 
than Japan and South Korea.
 Out of 56
countries in the APNIC region, the average annual growth rate contributed by these 5 countries is 47 $\%$.
\begin{table}
\caption{AS Count Long Term Trend  }
\label{table:AS_ltrend}       % Give a unique label
\begin{center}
\begin{tabular}{|c|c|c|c|c|c|}
\hline\noalign{\smallskip}
 Region / &Period &Model & Average   & Standard& Relative (APNIC)\\
 Country   &      &      & Annual  Growth   & Error & Growth Percentage\\
\noalign{\smallskip}\hline\noalign{\smallskip}
\hline
 APNIC&1994-2012& ARIMA(0,1,0)&464&64&100\\
      &         &Linear Model&453& 25&100\\
\hline
 India&1996-2012& ARIMA(0,1,0)& 38&9 & 8.1\\
      &         &Linear Model& 34 & 4 &7.5\\
\hline
 China& 1996-2012& ARIMA(0,1,0)&45& 9 &9.7\\
      &          & Linear Model&43& 3 & 9.5\\
\hline 
 Japan& 1995-2012 &ARIMA(0,1,0)&55& 8 &11.8\\
      &           &Linear Model&57& 2 & 12.6\\ 
\hline
 SKorea& 1995-2012 &ARIMA(0,1,0)& 59 &12&12.7\\
       &	   & Linear Model & 64 &3&14.1\\
\hline
Taiwan & 1995-2012 & ARIMA(0,1,0)&18&6&3.8\\
       &           & Linear Model&17& 1&3.7\\
   \noalign{\smallskip}\hline
 \end{tabular}
\end{center}
\end{table}
\\
\indent
It is observed that the long term trend curve computed using ARIMA model has structural difference between larger economies
and technology advanced countries.  Here we consider countries belong to larger economies as group I and countries
belong to technology advanced countries as group II. The structural deviation significance in the long term trend
between the two groups has to be established statistically. For this purpose correlation coefficients between
the series are analyzed. Correlation  \emph{$r_{1}$}  between long term trend 
series of group I countries i.e, India, China is computed. Similarly correlation \emph{$r_{2}$} is computed
 between long 
term trend series across group I and 
group II countries, i.e., India, Japan.
 Hypothesis testing on difference of sample correlation coefficients \emph{$r_{1}$} and  \emph{$r_{2}$} is used
  to determine the statistical significance of structural deviations.\\
\indent
\begin{table}
\caption{Correlations and Fisher Transformed Values}
\label{table:AS_correlations}       % Give a unique label
\begin{center}
\begin{tabular}{|c|c|c|c|}
\hline\noalign{\smallskip}
 Region / &Region/ & Corr. Coefficient & z-value   \\
 Country   & Country& r&    \\
\noalign{\smallskip}\hline\noalign{\smallskip}
\hline
 India& APNIC& .97 & 2.1\\
 \hline
 India&China& .98 & 2.1\\
 \hline 
 India& Japan& .86 & 1.3\\
 \hline 
 India& SKorea &.87 &1.3\\
 \hline
 India& Taiwan & .90 &1.5\\
 \noalign{\smallskip}\hline
 \end{tabular}
\end{center}
\end{table}
The population variables X and Y are AS count annual growth rate of group I and group II countries.
Bivariate normal distribution is assumed for the variables. The samples drawn for X and Y assign 
values of long term annual growth rate fitted from the ARIMA(0,1,0) model for a country. 
 The correlation coefficient r is computed between two samples drawn from the same group or across groups.
Fisher's transformation \citep{structural1} is applied on \emph{r} for variance stabilization. The procedure 
for hypothesis testing on significant statistical difference between correlation coefficients is as 
follows:
\begin{enumerate}
 \item r1 is correlation coefficient computed between  samples  of  size n1 within group I.
 \item r2 is correlation coefficient computed between samples of size n2 across group I and II .
 \item z1 and z2 are computed using Fisher's transformation applied on r1 and r2 for variance stabilization.
 \item z = $\frac{1}{2*ln[\frac{(1+r)}{(1-r)}]}$
 \item zd = $\frac{(z1-z2)}{\sqrt{(\frac{1}{(n1-3)}+\frac{1}{(n2-3)})}}$
 \item The difference zd is assumed to be standard normal
 \item H0: z1 and z2 are equal, H1: z1 and z2 are different
 \item If absolute value of zd is less than 1.96 then accept null hypothesis H0
 \item Otherwise reject null hypothesis with 95 $\%$ confidence level
 \item p-value is computed as pnorm(zd)*2 when zd is negative
 \item Otherwise p-value is computed as (1-pnorm(zd))*2 
\end{enumerate}
The r and z values computed with and across groups are given in table \ref{table:AS_correlations}. We can observe that
for India r value is high within the group and less across the groups. This can be interpreted as India has strong
structural similarities in long term trend pattern with in group I countries and 
significant structural difference with Group II countries.
 The statistical significance of difference in z1 
and z2 and p-values for India are given in table \ref{table:AS_corrdiff}. The zd values show that the long term trend
is significantly different between India and the Group II countries.
\begin{table}
 \caption{Correlation Difference and p-value for India}
\label{table:AS_corrdiff}       % Give a unique label
\begin{tabular}{|c|c|c|c|c|c|}
\hline\noalign{\smallskip}
 With in  & z1 - value & Across Group I & z2-value & zd  & p-value   \\
 Group I   &          &  and II        &         & (difference)    &    \\
\noalign{\smallskip}\hline\noalign{\smallskip}
\hline
 India and China &  2.1    &   India and Japan & 1.3     & 2.11 & .03\\
 \hline 
 India and China&  2.1    &   India and Taiwan & 1.5   & 1.66 & 0.09\\
  \noalign{\smallskip}\hline
 \end{tabular}
\end{table}
\section{Inter Annual Variations}
\label{sec:4}
\begin{figure}
 \centering
  \includegraphics[width=0.4\textwidth]{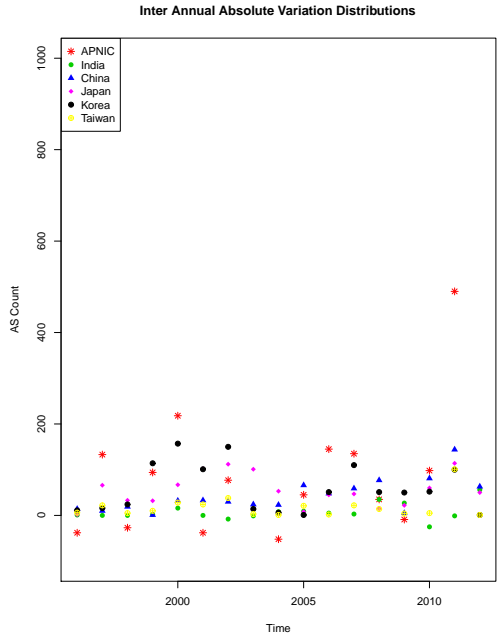}
  \caption{AS Count IAAV Distribution Data }
 \label{fig:AS_IAAV}
\end{figure}

The Inter Annual Variation (IAV) is another component of the time series. These values are random
and does not contribute to the long term trend. Each country has different number of AS registrations
 every year depending upon the demand within 
the country. The demand for new ASes each year is influenced by the factors such as
 increasing number of new ISPs, content providers, application providers and future reservations. All these
factors can be represented using the macro economic variable service sector industries. If there
is significant AS count variation in a year, it should have occurred due to the domino effect of
fluctuations occurred in some of the aforementioned factors.
\\
\indent
The IAV values are either positive or negative. We consider the absolute values for our analysis.
It is denoted by the term Inter Annual Absolute Variation (IAAV). The computation of it is carried out
 as $Y_{t} - Y_{t-1}$, which is the first differenced 
data of the time series.
 The year wise AS count data has an additive trend. But IAAV is computed on the first difference of
 the year wise data. To check for the presence of trend in the IAAV, ACF and  PACF values are tested. Also
normality hypothesis test is performed on the values to fix the distribution.
The ACF and PACF values are non significant for all the lags of the computed 
IAAV data to the considered countries excluding India. In order to remove any trend component 
present in the data for India we have computed the second differencing. 
In these series, data are independent. We have done the statistical tests Shapiro-Wilk and Jarque Bera to find the 
presence of normal distribution in the IAAV data. Both the tests assume normal distribution
for the data as null hypothesis. The p-values of the tests are greater than .05 only for Japan, 
which confirms the acceptance of null hypothesis.
For other countries the p-values are non significant. Hence normality assumption to the data is rejected.\\
\indent
To detect the significant event i.e., change point in variances, we assume the data are independent
  and the distribution of the data is unknown. The IAAV distributions for the countries are 
shown in Figure \ref{fig:AS_IAAV}.
\indent 
\subsection{Event Detection}
\begin{figure*}
 \centering
 %yy\includegraphics[width=0.75\textwidth]{./graphs/as_IAVP_Trend_linear_ARIMA_india.eps}
 \includegraphics[width=0.5\textwidth]{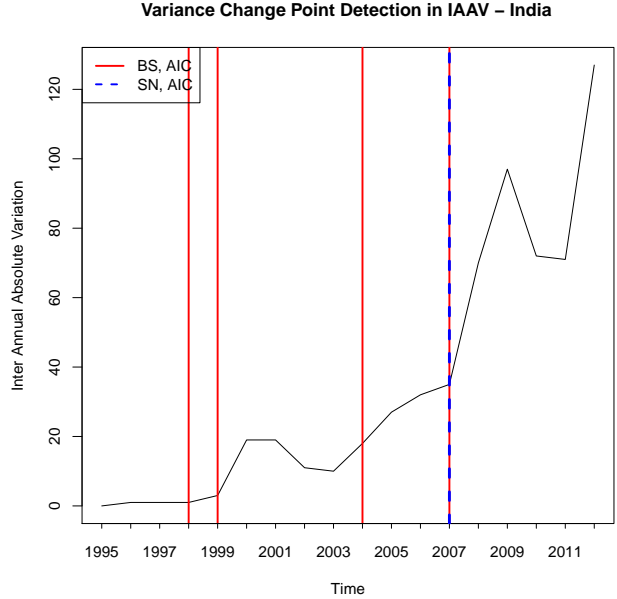}
\caption{Change Point Detection Using AIC Penalty}
 \label{fig:varcpdASI1}
\end{figure*}

\begin{figure*}
 \centering
 %yy\includegraphics[width=0.75\textwidth]{./graphs/as_IAVP_Trend_linear_ARIMA_india.eps}
 \includegraphics[width=0.5\textwidth]{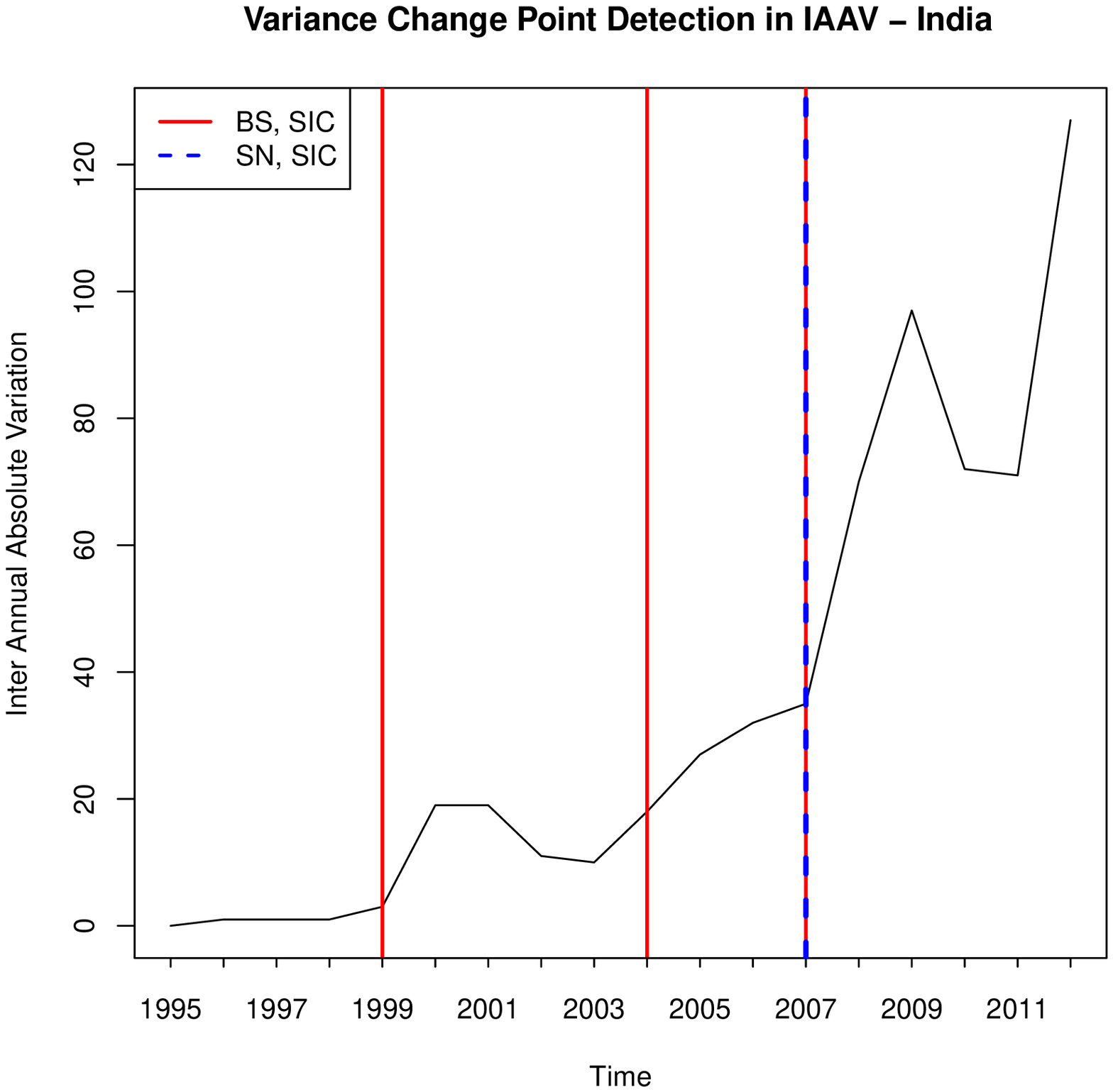}
\caption{Change Point Detection Using SIC Penalty}
 \label{fig:varcpdASI2}
\end{figure*}

 We define  significant variation in IAAV as an event. The impact of the factors
on the significant events of the data are ill understood or unpredictable in reality. So we use the statistical
properties of the sample data for detecting events \citep{eventdetection1}.
 This is considered as Change Point Detection (CPD) problem in statistics. \\
\indent
We are using batch mode CPD for identifying the significant changes in IAAV.
 In batch mode, all the change points are detected at once. Since the distribution of the data is assumed 
to be unknown,  CUSUM \citep{cusum} is used as test statistic for CPD.
 The CUSUM method followed in \citep{cusum} is used in our
work and given in 
algorithm \ref{algo:cusum}.
Using the computed CUSUM test statistic, the Binary Segmentation (BS) and 
Segment Neighborhood (SN) methods are adapted to search for multiple change points.
The BinSeg algorithm \citep{binsegment} uses
single change point detection method initially on the entire series to detect \emph{t} (time) satisfying the equation
\ref{equ:costpenal1}. If the condition is true then segmentation is performed on the identified \emph{t}. 
On the two new segments, the procedure is iterated until no other change point is detected.
  It is
an approximate method with computational cost as $\BigO{nlogn}$ where n is number of elements present in the time series.
\begin{equation}
\label{equ:costpenal1}
 C(Y_{1:t}) + C(Y_{t+1:n}) + \beta  < C(Y_{1:n})
\end{equation}
\begin{algorithm}
\caption{CUSUM Algorithm}
\label{algo:cusum}
\begin{algorithmic}
%\Comment{IAAV - Inter Annual Absolute Variance Data }
\STATE $DATA \Leftarrow IAAV $ 
\STATE $CUSUM \Leftarrow 0$
\STATE $m = mean(DATA) $
\STATE $len = length(DATA)$
\STATE $con = 0 $
%\While{$con \lt len $}
\REPEAT
\STATE  $CUSUM{_i}$ = $CUSUM{_i}$ +  ($DATA{_i}$ - m), i=1 \dots len
\STATE $con = con+1$
\UNTIL{$con \le len $}

\end{algorithmic}

\end{algorithm}
\indent
The SN algorithm \citep{segneigh} uses exact search method based on dynamic programming. 
Number of change points we like to search for can be specified with parameter Q.
 It is assumed as the upper limit on the
number of segments.  The cost function is computed for all possible segments in the entire series.
Change points between 0 to Q are considered from all the possible segments. In addition to the cost function,
a penalty value can also be incorporated in the search method.   
As a consequence of exhaustive search, the computational cost is $\BigO{Qn{^2}}$ for the algorithm.
 When the number of change points increases linearly
 with time, the computational cost will increase cubic in the size of the series. Both of the search methods use a common
approach based on minimizing cost and a penalty function using equation \ref{equ:costpenal} to perform the segmentation \citep{PELTcpd}.
\begin{equation}
 \displaystyle\sum\limits_{i=1}^{m+1}[C(Y(t_{i-1}+1):t_{i})] + \beta f(m)
\label{equ:costpenal}
\end{equation}
Here C is the cost function of the time series segment $Y(t_{i-1}+1):t_{i}$ and $\beta f(m)$ is
 the penalty function to protect against over fitting. In the change point detection literature \citep{varcpd},
 twice the negative log likelihood is used commonly as cost function. The penalty function is generally
used as linear in the number of change points \emph{m} i.e., $\beta f(m)= \beta m$.
 Akaike Information Criteria (AIC) \citep{aic} and Schwarz
Information Criterion (SIC) \citep{bic}  are such penalty functions. AIC  uses  $\beta = 2p$ and SIC uses
 $\beta = plog(n)$ as penalty values. In this \emph{p} is the number of additional parameters introduced by adding a
change point and \emph{n} is number of elements in the series. The SN algorithm guarantees the global minimum of equation
\ref{equ:costpenal} but BS does not provide the guarantee.
\\
\indent
We have used AIC and SIC as penalty functions and find the variance change point in IAAV. During estimation
of CPD, we have observed multiple smaller segments due to small penalty value of AIC and local minima of BS.
 In such conditions, we consider the change point commonly detected by both the methods with SIC as penalty function. 
Change point detection in IAAV level change for India is given in Figures \ref{fig:varcpdASI1} and \ref{fig:varcpdASI2}  . 
\\
\indent
 We have observed that the BS method detects four change points in the variance level and SN detects one change point
when AIC is used as penalty function. When SIC is used as a penalty function, BS detects three change points and SN detects
one change point. The change point detected in the year 2007 for the level variation is common to both the methods. 
The graphical analysis of the IAAV graph also confirms this level change in variation.
 Hence we can conclude that
there is a significant level change in variation occurred during 2007 for  India. We have observed significant
 level change in IAAV variation in 2004 for China by both the search methods. But for IAAV series of Japan,
 there is no level change observed in variance
 by both the methods. In case of South Korea, both the methods detected three change points when AIC is used as a penalty
function. But no change point is detected when SIC is used as penalty function. Since
the segment size performed by both the search methods are small when AIC is used, we consider the
result of SIC as penalty function. In case of Taiwan, SN and BS detect the change point
in 2010 for both the penalty functions. The whole APNIC region has a change point for IAAV variations level in
2006. This change point for APNIC region is detected by using SIC as penalty function. The  change point year, and the
estimated yearly growth rate of IAAV before and after the change point are given in table  \ref{table:varcpdAS}.
\begin{table}
\caption{Change Point Detection Details for IAAV}
\label{table:varcpdAS}       % Give a unique label
\begin{center}
\begin{tabular}{|c|c|c|}
\hline\noalign{\smallskip}
 Country/Region & Change Point  & IAAV Growth rate (\%) \\
                &               & (before, after) cpd \\ 
\noalign{\smallskip}\hline\noalign{\smallskip}
\hline
 APNIC & 2006 & 0.33,1.54\\
 India & 2007 & 0.47,2.3 \\
 China & 2004 & 0.34,-0.05\\    
 Japan & No change& -0.08\\
 SKorea & No change& -.03\\
 Taiwan & 2010 & .02,-32.4\\
\noalign{\smallskip}\hline
 \end{tabular}
\end{center}
\end{table}
\\
\indent
We can infer that the yearly change in variation is with positive rate of 2.3 $\%$ for India after 2007.
 The variation is constant for countries in advanced technology countries from 1995 to 2012 except an outlier detected to
Taiwan during the year 2010. After 2004 the variation is constant for China also. India and the APNIC region have significant
percentage of change in IAAV after the change point detection. We analyze the macro economic variable service sector
growth which is an aggregate measure of factors that influence the IAAV to
 understand the significant change in variations for India. The Gross Domestic Product (GDP) is used as a proxy
variable for service sector growth.
\subsection{IAAV Events and GDP}
We have observed significant changes in IAAV after 2007. To understand the driving force for these variations  
we analyzed the GDP of India. GDP is a measure
of yearly output products and services of a country. Indian GDP is a composition of outputs from agriculture, industry
and services. After the post economic reform period (from 1991 onwards) GDP is mainly driven by the services sector. The
table \ref{table:GDP_composition} shows the contribution of services sector to GDP during the years. The data is
taken from economic survey \citep{economicsurvey1} and various publications of Reserve Bank of India (RBI). The percentage
share of service sector is steadily increasing from 1991 onwards and more than 64 $\%$ during 2008-2009. Hence we
use GDP as a proxy variable for services sector growth.
\begin{table}
\caption{The Components Percentage Share to GDP}
\label{table:GDP_composition}       % Give a unique label
\begin{center}
\begin{tabular}{|c|c|c|c|}
\hline\noalign{\smallskip}
 Year & Agriculture  & Industry  & Services \\
 \noalign{\smallskip}\hline\noalign{\smallskip}
\hline
 1990-91 & 31.4 & 19.8 & 48.8\\
 1995-96 & 27.3 & 21.2 & 51.4 \\
 2000-01 & 23.9 & 20.4 & 56.1 \\    
 2005-06 & 19.5 & 19.4 & 61.1\\
 2008-09 & 17.0 & 18.5 & 64.5\\
 2012-2013& 14.1& 21.1 & 64.8\\
 \noalign{\smallskip}\hline
 \end{tabular}
\end{center}
\end{table}
\\
\indent
The annual growth rate of GDP is taken from economic survey and given in table \ref{table:gdp}. During the years
2006-2008, the GDP witnessed a growth rate of more than 9 $\%$ driven by contributions from services sectors such as telecommunication, 
computer software, railways and education. IAAV change point significantly correlates with this time period. This
correlation establishes that the services sector industry growth influences the variations in IAAV significantly. 
 \begin{table}
\caption{GDP during 1992 -2010 }
\label{table:gdp}       % Give a unique label
\begin{center}
\begin{tabular}{|c|c|}
\hline\noalign{\smallskip}
 Year & Annual GDP \\
      & Growth Rate \\
 \noalign{\smallskip}\hline\noalign{\smallskip}
\hline
 1992-93 & 5.4 \\    
 1993-94 & 5.7 \\
 1994-95 & 6.4 \\
 1995-96 & 7.3 \\
 1996-97 & 8.0 \\
1997-98 & 4.3 \\
1998-99 & 6.7 \\
1999-00 & 6.4\\
2000-01 & 4.4 \\
2001-02 & 5.8 \\
2002-03 & 3.8 \\
2003-04 & 8.5 \\
2004-05 & 7.5 \\
2005-06 & 9.5 \\
2006-07 & 9.7 \\
2007-08 & 9.0 \\
2008-09 & 6.7 \\
2009-10 & 7.2 \\
 \noalign{\smallskip}\hline
 \end{tabular}
\end{center}
\end{table}
\section{AS Routeview Data Analysis}
\label{sec:5}
\begin{figure*}
 \centering
 %yy\includegraphics[width=0.75\textwidth]{./graphs/as_IAVP_Trend_linear_ARIMA_india.eps}
 \includegraphics[width=0.5\textwidth]{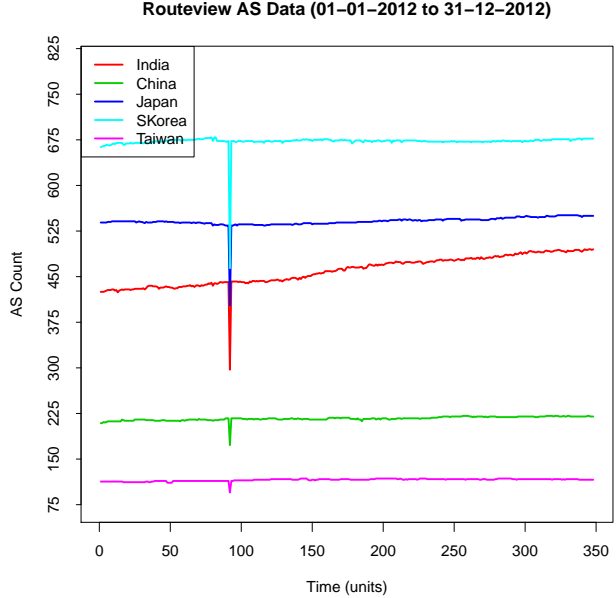}
\caption{AS Time Series Data Estimated from Routeview}
 \label{fig:asRview}
\end{figure*}

To understand the temporal presence of  assigned ASes of each country in the global routing table,
we have analyzed the routeview data \citep{routeview} from 01-01-2012 to 01-01-2013. The sample snapshot is taken
at fixed 0000 hours of the day. We consider the per day AS count data for the analysis. This AS count is obtained
by searching for registered AS numbers in
APNIC registry of each country in the routeview daily snapshot. The presence of AS numbers in the routing table can be 
interpreted as currently advertising ASes from the assigned ASes of a country. 
These ASes can be reached from  one or more geographic regions using the E-BGP paths. At least
one prefix of the AS is being advertised by the AS. The AS may be originating AS, transit AS or both. 
Events such as link failures, node failure and attacks will cause reachability problems to the ASes. The impact of the
events will be manifested in the form of stochastic variations in AS counts estimated from the global routing table.   
The estimated AS data for the countries under consideration are given in Figure \ref{fig:asRview}.
We can observe a clear  linear growth of AS count in the routeview table during 2012-2013 for India. 
The increase is estimated as 16 $\%$.
for this period. For the same period, the rest of the countries have growth  below 6 $\%$. A significant 
stochastic variation is also observed on $17{^th}$ April 2012. During this time period 35 $\%$ of ASes are
not reachable for India, Japan and South Korea whereas 25 $\%$ of ASes are not reachable for China and Taiwan. 
This may be due to a failure in common link that is used to reach ASes of these countries and the reachability is
restored within a day. The advertised Vs assigned ratio is computed using the APNIC assigned AS data and routeview
AS data. When we consider the whole APNIC region, only 60 $\%$ of the assigned ASes are advertising in the global
routing table. It can be interpreted as  only 60 $\%$  of the assigned ASes are reachable from different geographic
locations. For China,  40 $\%$ of the ASes can only be reachable from global sources. This ratio from the data 
indicates that 20 to 60 $\%$ of the assigned ASes are not reachable from outside to the countries in the APNIC region.
 This may be due to the reason
that the AS may not be operational or used only in I-BGP. The advertised Vs assigned AS details are given in table
\ref{table:asAssiADV}.
 \begin{table}
\caption{Assigned Vs Advertised AS Count 2013}
\label{table:asAssiADV}       % Give a unique label
\begin{center}
\begin{tabular}{|c|c|c|c|c|c|}
\hline\noalign{\smallskip}
 Country/Region & Registered & Assigned & Advertised & Ratio& Increase $\%$\\
                &            &          &            &      & (2012-13)\\
\noalign{\smallskip}\hline\noalign{\smallskip}
\hline
 APNIC & 9876&8420 & 5285 & .6 & -\\
 India & 614 & 607 & 495  & .8 &16.4\\
 China &729 & 551 & 220 & 0.4& 5.2 \\    
 Japan & 993& 800 & 550 & 0.7& 2.04 \\
 SKorea & 1016& 857& 677 &0.8& 2.11 \\
 Taiwan & 308 & 196& 116 &0.6& 2.65 \\
\noalign{\smallskip}\hline
 \end{tabular}
\end{center}
\end{table}
\\
\indent
The analysis of AS data in routeview table shows the highest increase of 16 $\%$ in the AS global reachability for
India during the period 2012-2013.  In the assigned ASes, 80 $\%$ of the ASes can be reached from various 
global locations. The stochastic variations of short durations
in AS count time series are more likely in the routeview data due to the occurence of various external events such
as attacks and under sea cable cuts. The study on
these stochastic variations and inferring the events that caused variations are our future work.  
\section{Related Work}
\label{sec:9}
So far in the Internet AS topology Research domain, structural analysis on the AS topology \citep{topostruct1},
 topology generation methods \citep{topogen1}
and impact of routing dynamics on AS topology \citep{toporoute1,toporoute2} are studied.
 There are different topological features such as AS node count, 
average node degree, the node degree relationship with node count and different centrality measures
 are reported in AS structural studies. Prefix counts, 
path distributions, average path length and peer counts are reported in AS routing dynamic studies.
 Sites like Potaroo.net \citep{asn32} and hurricane electric \citep{he}  provides daily reports on ASes, prefixes, peers, routing table size,
 withdrawn, newly announced  and bogus routes. Weekly, monthly and yearly summaries are also provided by these sites
 on the aforesaid features for country wise. Still understanding on the temporal occurrence of events specific to a country, 
impact of the events on the trend and yearly variances of the AS topology are to be explored completely. In this
 work we have chosen the AS node count to explore its growth and variations by events internal and external to a country.
\\
\indent
The growth trends in the number of ASes seen in the global routing system and the RIRs are studied 
by Dhamdhere et al.\citep{asmodel1, asmodel2}. The main observation they have made in their study is
 that ARIN and RIPE RIRs have
shown distinctly different growth trends since 2001 in terms of the number of advertised ASes. Until mid 2001, 
both RIRs showed exponential growth trend. After that ARIN
has grown linearly and RIPE changed to a slower but exponential increase. The number of advertised ASes is larger
in RIPE than ARIN. 
\\
\indent
The AS number resource consumption is extensively studied and prediction for the pool exhaustion is
performed by Geoff Houston in his work \citep{ASN}. In the unstructured 16 bit AS numbers, excluding
the reserved numbers like 0, 65,535 and private pool from 64512 through 65534, effectively 1 through 64511
AS numbers are available for global Internet routing. The ASN consumption per year in the global pool
is reported as 3500 from 2002 onwards. This is roughly 5.4 $\%$ per year. He also reported the RIR pool size
in 2006, advertised  and assigned ASes. The prediction models he used are exponential and linear.
In the exponential model, recent past 3 year values are considered for future predictions.
\\
\indent
In our work we have considered time series ARIMA models that use past values and errors in forecasting. The
statistical properties of the long term trend for technology advanced countries and large economies are analyzed.
We also use change point detection methods on Inter Annual Absolute Variation to detect abrupt changes in the
growth within a country. Our work is oriented towards AS resource planning,  policy making and growth anomaly
detection with respect to a country.
\section{Conclusion}
\label{sec:7}
In this work we have analyzed the AS resource data available in the APNIC repository for five Asian countries
and the APNIC region.
The countries are chosen based on two categories namely fast growing economies and technologically advanced
 countries in the APNIC region. The characterization on the AS count time series data is performed to identify the
appropriate time series model. The estimated autocorrelation properties upto lag 3 and partial autocorrelation
properties upto lag 1  indicate the presence of AR and MA components. The location change in the data indicates
a linear trend. Based on these characterization ARIMA models are chosen to forecast the data. The chosen model validation is performed
 by analyzing the residuals for randomness, constant variations and normal distribution.
 From the model candidates, forecasting accuracy is used as an important criteria 
in final model selection. An out of sample forecasting with point forecasts, prediction intervals and prediction
accuracy are reported for the selected model of each country. The long term trend is analyzed using ARIMA(0,1,0)
and Linear trend models. The average yearly growth rate and the trend direction are reported. The hypothesis
test establishes significant structural deviations in long term trend of India and the technology advanced countries.
 From the IAAV
data, we analyzed for change in variations with CUSUM test statistic using BS and SN search methods.
 Two penalty functions AIC and BIC are used
along with the search methods. The level change in variations and yearly growth rate are reported. The change
in IAAV at a rate of 2.3 $\%$ after 2007 is hypothesized as driven by growth in Indian services sector.
This is established using the evidence from GDP data assumed as proxy variable for services sector. The technology
 advanced countries witness a constant IAAV value during the observation period. The global
reachability  percentage during 2013 with respect to assigned ASes and  the yearly growth percentage
 are reported. The significant level change
in variations, positive growth percentage in IAAV and higher percentage of advertised ASes when compared 
to other countries indicate India's fast growth  and wider global reachability of Internet infrastructure from 
2007 onwards. Our modeling effort reveals new insights and patterns in the country level Internet infrastructure 
indicator: 
AS count, for the countries that fall under two groups. The stochastic variation analysis in the global AS reachability
data and event detection are considered for our future work.

%\begin{acknowledgements}
%\end{acknowledgements}

\bibliographystyle{spbasic}      % basic style, author-year citations
\bibliography{statistical_analysis}   % name your BibTeX data base

%\begin{thebibliography}{}
% etc
%\end{thebibliography}

\end{document}